\documentstyle[twoside,fleqn,espcrc2,epsfig]{article}

\def\bold#1{\setbox0=\hbox{$#1$}%
     \kern-.025em\copy0\kern-\wd0
     \kern.05em\copy0\kern-\wd0
     \kern-.025em\raise.0433em\box0 }

\def\ga{\mathrel{\raise.3ex\hbox{$>$\kern-.75em\lower1ex\hbox{$\sim$}}}}
\def\la{\mathrel{\raise.3ex\hbox{$<$\kern-.75em\lower1ex\hbox{$\sim$}}}}

\def\m12{m_{1\!/2}}

\begin{document}
\title{Neutrinos from the Annihilation or Decay of Superheavy Relic Dark Matter Particles}

\author{Dan Hooper and Francis Halzen
\address{University of Wisconsin, 
1150 University Ave., Madison, WI~53706, USA}}

\begin{abstract}
In light of the mounting evidence that the highest energy cosmic rays are dominated by protons and not gamma-rays, we discuss the prospect that these cosmic rays are generated in the decay or annihilation of superheavy relic particles.  We calculate the high energy neutrino spectrum which results and normalize our results to the ultra-high energy cosmic ray spectrum.  We show that most scenerios are already constrained by present limits placed by the AMANDA experiment.
\end{abstract}

\maketitle

\section{Introduction}

The discovery of cosmic rays with energy exceeding the GZK cutoff
\cite{gzk} presents an interesting challenge to astrophysics, particle
physics, or both.  Numerous scenarios have been
proposed to solve the problem. These include exotic particles
\cite{exo}, neutrinos with QCD scale cross sections
\cite{qcdneu}, semi-local astrophysical sources
\cite{loc} and top-down models \cite{top}.

Recent measurements confirm that our universe contains a large fraction
of cold
dark matter \cite{cmb}.  A top-down model in which annihilating or
decaying superheavy particles produce the highest energy cosmic
rays could potentially solve both of these problems \cite{wim1,wim}.

Conventional particle physics implies that ultra high-energy jets
fragment predominantly into photons with a small admixture of protons.  This seems to be in disagreement with mounting evidence
that the highest energy cosmic rays are not photons \cite{pro1}. This does not
necessarily rule out superheavy particles as the source of the highest
energy cosmic rays.  The uncertainties associated with the cascading of
the jets in the universal radio background and with the strength of
intergalactic magnetic fields leave open the possibility that ultra
high-energy photons may be depleted from the cosmic ray spectrum near
$10^{20}$ eV, leaving a dominant proton component at GZK energies
\cite{radio}.  With this in mind, we will choose to normalize the proton
spectrum from top-down scenarios with the observed ultra high-energy
cosmic ray flux.

Neutrinos are produced more numerously than protons and travel much
greater distances. The main point of this note is to point out that
this ``renormalization'' of the observed cosmic ray flux to protons
generically predicts observable neutrino signals in operating
high-energy neutrino telescopes such as AMANDA.

\section{Nucleons from Ultra High-Energy Jets}

To normalize the production rate of ultra high-energy jets,
it is necessary to calculate the
spectrum of nucleons resulting from their fragmentation.  Each
jet will fragment into a large number of hadrons approximated by a
fragmentation function rooted in accelerator data \cite{frag}. All hadrons produced eventually decay into pions and nucleons.  For a detailed discussion of ultra high-energy fragmentation, see Ref.\,\cite{frag2}.

To solve the ultra high-energy cosmic ray problem, this nucleon flux must
accommodate the events above the GZK cutoff.  Observations
indicate on the order of $10^{-27}$ events $\rm{m}^{-2} \rm{s}^{-1}
\rm{sr}^{-1} \rm{GeV}^{-1}$ in the energy range above the GZK cutoff ($5
\times 10^{19}$ eV to $2 \times 10^{20}$ eV)\cite{agasa}. The
formalism of a generic top-down scenario is sufficiently flexible to
explain the data from either the HIRES or AGASA experiments. For an example, see Figure 1.

\begin{figure}[h] 
\vbox{\kern2.4in\special{ 
psfile='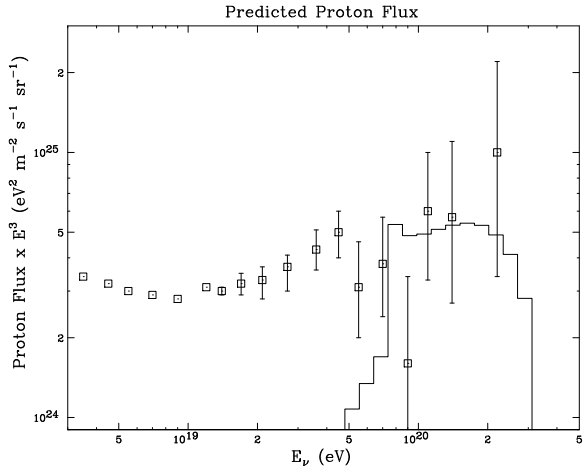' 
angle=0 
voffset=-90
hoffset=-30
hscale=45 
vscale=45}} 
\vspace*{-5mm}
\caption{The ultra high-energy cosmic ray flux predicted from the
decay or annihilation of superheavy dark matter particles producing
$10^{21}$ eV
hadronic jets is compared to the AGASA cosmic ray
data.  The distribution of dark matter used is isotropic with an
overdensity factor of $10^5$ within 20 kpc.  Note
that all
observed super GZK events can be explained by this mechanism.}
\vspace*{-5mm}
\end{figure} 

\vspace{10mm}

\section{Cross Sections and Lifetimes for Superheavy Particles as Dark Matter}

If the superheavy relics responsible for the highest energy cosmic rays are also the  
solution to the dark matter problem, then 
\begin{eqnarray}
\rho_X \sim 0.3 \times \rho_{\rm{critical}} \simeq 0.3 \times 8.45  
\times 10^{-27} \rm{kg}/\rm{m}^3,
\end{eqnarray}
and may be much larger locally.  Therefore, the lifetime of such a decaying  
particle must be:
\begin{eqnarray}
\tau_X \sim \frac{\rho_X}{m_X \frac{dn_X}{dt}} \sim  
10^{17} \rm{years}.
\end{eqnarray}
Such a long lifetime may be disfavored by fine-tuning arguments, but can be possible \cite{wim1,life}.  If, however, the  
superheavy relics in question were not the major dark matter component,  
this lifetime could be much shorter.

If instead we consider stable particles which can annihilate with each  
other, we can calculate their annihilation cross section as a function  
of velocity.  For an isotropic distribution of particles,
\begin{eqnarray}
\frac{\rho_X^2}{m_X^2} \sigma_{XX} v_{\rm{rms}} \simeq \frac{dn_X}{dt}.
\end{eqnarray}
For a extragalactic distribution,
\begin{eqnarray}
\sigma_{XX} \sim \frac{2 \times 10^{-15} \rm{m}^3  
\rm{s}^{-1}}{v_{\rm{rms}}},
\end{eqnarray}
or, for a galactic halo distribution,
\begin{eqnarray}
\sigma_{XX} \sim \frac{10^{-19} \rm{m}^3  
\rm{s}^{-1}}{v_{\rm{rms}}}.
\end{eqnarray}
A characteristic velocity of 500 km/s, for example, would correspond to a annihilation  
cross section of $\sim10^7$ bn for extragalactic dark matter or $\sim 10^{3}$ bn for a galactic halo distribution.  If the dark matter were not uniformly  
distributed, however, but were distributed in clumps with  
characteristic densities of $\rho_{\rm{clump}}\sim C \rho_{\rm{mean}}$  
then the annihilation cross section would be lowered by a factor of  
$C$.  For example, if the density of a typical cluster were $10^6$  
times greater than in an isotropic distribution, annihilation cross  
sections could be in the mb range.  It is interesting to note that if superheavy dark matter is distributed locally with mb elastic scattering cross sections, they would become gravitationally trapped in the sun and annihilate as described in Ref.\,\cite{sim}.  

\section{Neutrinos from Ultra High-Energy Jets}

Neutrinos are produced in several ways in the fragmentation of ultra high-energy jets: in semileptonic bottom and charm decays, in W producing top decays and most importantly, in the decay of charged pions. For a detailed discussion, see Ref.\,\cite{spec2}.

To obtain the neutrino flux, we multiply the injection spectrum by the
average distance traveled by a neutrino and by the rate per volume for
hadronic jets which we calculated earlier.  Neutrinos, not being
limited by scattering, travel up to the age of the universe at the
speed of light ($\sim$ 3000 Mpc in an Euclidean approximation). The predicted neutrino flux is shown in Figure 2.  Note that this
flux is significantly higher than the present limits placed by the
AMANDA experiment\cite{Andres:2001ty} for the case of an isotropic
distribution of ultra high-energy jets.  For the galactic
scenario, however, the flux predicted is comparable with present
AMANDA-B10 limits.  Also shown are the limits anticipated from AMANDA-II
data and the IceCube experiment.

\begin{figure}[thb] 
\vbox{\kern2.4in\special{ 
psfile='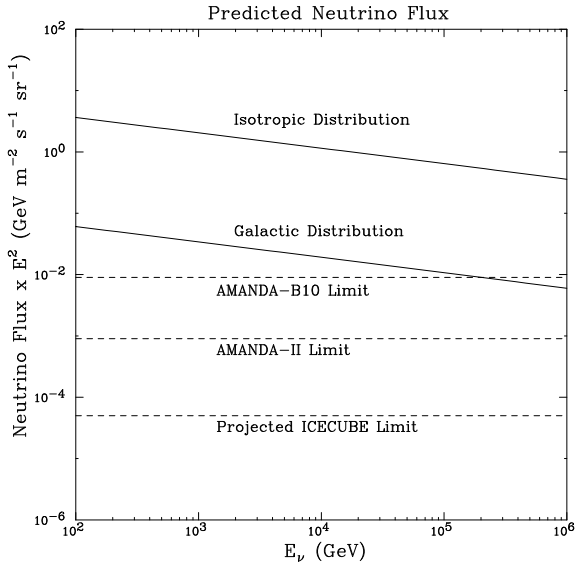' 
angle=0 
voffset=-100
hoffset=-30
hscale=45 
vscale=45}} 
\vspace*{-2mm}
\caption{The muon neutrino flux from $10^{21}$ eV
jets normalized to the highest energy cosmic rays. Shown are the fluxes
calculated for an isotropic and for a galactic distribution of jets.  The
dotted lines shown are the experimental diffuse flux limits from
present AMANDA-B10 data, projected AMANDA-II data and projected
IceCube data.  These limits are $E^2 \frac{dN}{dE} \le 9\times 10^{-7}$,
$9\times 10^{-8}$ and $5\times 10^{-9}$ GeV $\rm{cm}^{-2} \rm{s}^{-1}
\rm{sr}^{-1}$ respectively.  Note that operating experiments can
effectively test this range of models.}
\end{figure} 

\vspace*{-10mm}

\section{Event Rates in High-energy Neutrino Telescopes}

The diffuse flux of high-energy neutrinos can be observed by operating
and planned neutrino telescopes.  AMANDA-B10, with an effective area of
$\sim$5,000 square meters has placed the strongest limits on the diffuse
neutrino flux.  Figure 2 shows that for a galactic distribution of
superheavy dark matter particles, the predicted high-energy neutrino
flux is on the order of the present diffuse flux limit.  The neutrino
flux for an isotropic distribution of ultra high-energy jets is well
above the present limits.  If superheavy particles are indeed the source of the highest energy cosmic rays,
observations by high-energy neutrino telescopes should reveal the
corresponding neutrino flux. For a review of cerenkov neutrino
telescopes see Ref.\, \cite{Andres:2001ty,ice2}. 

Note that gamma ray astronomy can also provide an interesting test of these scenarios \cite{gamma}.

\section{Conclusions}

If the decay or annihilation of superheavy relics is the source of the highest energy cosmic rays, then a
high-energy neutrino flux should
accompany the observed cosmic ray flux.  This neutrino flux will be
much higher than the flux of nucleons due to the much greater mean free
path of neutrinos and greater multiplicity of neutrinos produced in
high-energy hadronic jets.

The high-energy neutrino flux generated in such a scenario can be
calculated by normalizing the flux of appropriate particles to the ultra
high-energy cosmic ray flux.  With mounting evidence that the highest
energy cosmic rays are protons or nuclei and not photons, we must
require that the ultra high-energy photons are degraded by the universal
radio background, leaving protons to dominate the highest energy cosmic
ray flux.  The neutrino flux must be normalized to the proton flux
resulting in significantly improved prospects for its detection.

This paper shows that the neutrino flux accompanying the highest energy
cosmic rays in these models is on the order of the limits placed by
operating high-energy neutrino telescopes such as AMANDA.  Further data
from AMANDA II, or next generation neutrino telescope IceCube, can test
the viability of models in which superheavy particles generate the highest energy
cosmic rays.

\section{Acknowledgements}
We would like to thank John Beacom, Bob McElrath and Michael Kachelriess for valuable discussions. This work was supported in part by  
a DOE grant No. DE-FG02-95ER40896 and in part by the Wisconsin Alumni  
Research Foundation.


\begin{thebibliography}{9}





\bibitem{gzk}
K.~Greisen,
Phys.\ Rev.\ Lett.\  {\bf 16}, 748 (1966); G.~T.~Zatsepin and
V.~A.~Kuzmin,
JETP Lett.\  {\bf 4}, 78 (1966).




\bibitem{exo}
A.~Perez-Lorenzana, {\it Oaxaca de Juarez 1998, Particles and Fields 409-412};
L.~Masperi,
{\it In Trieste 1998, Non-accelerator particle astrophysics 218-224};
I.~F.~Albuquerque, G.~R.~Farrar and E.~W.~Kolb,
Phys.\ Rev.\ D {\bf 59}, 015021 (1999),
hep-ph/9805288.


\bibitem{qcdneu}
A.~Jain, {\it et al.}, hep-ph/0011310; P.~Jain, {\it et al.}, Phys.~Lett.~B {\bf 484}, 267 (2000), hep-ph/0001031; G.~Domokos, {\it et al.}, hep-ph/0006328; M.~Kachelriess and M.~Plumacher, Phys.~Rev.~D {\bf 62}, 103006 (2000), astro-ph/0005309.



\bibitem{loc}
C.~Isola, {\it et al.}, astro-ph/0104289; P.~Blasi and A.~V.~Olinto,
Phys.\ Rev.\ D {\bf 59}, 023001 (1999), astro-ph/9806264.




\bibitem{top}
R.~J.~Protheroe and T.~Stanev,
Phys.\ Rev.\ Lett.\  {\bf 77}, 3708 (1996), astro-ph/9605036; P.~Bhattacharjee, {\it et al.}, Phys.\ Rev.\ Lett.\  {\bf 69}, 567 (1992); T.~J.~Weiler,
Astropart.~Phys.~{\bf 11}, 303 (1999); S.~W.~Hawking,
Nature {\bf 248}, 30 (1974); F.~Halzen, {\it et al.}, Phys.\ Rev.\ D {\bf 52}, 3239 (1995), hep-ph/9502268.




\bibitem{cmb}
C.~B.~Netterfield {\it et al.},
astro-ph/0104460; C.~Pryke, {\it et al.}, astro-ph/0104490; A.~Balbi {\it et al.},
Astrophys.\ J.\  {\bf 545}, L1 (2000), astro-ph/0005124;
S.~Perlmutter {\it et al.}  [Supernova Cosmology Project Collaboration],
Astrophys.\ J.\  {\bf 517}, 565 (1999), astro-ph/9812133.




\bibitem{wim1}
E.~W.~Kolb, {\it et al.}, hep-ph/9810361.

\bibitem{wim}
D.~J.~Chung, {\it et al.},
Phys.\ Rev.\ D {\bf 59}, 023501 (1999), hep-ph/9802238;
P.~Blasi, {\it et al.}, astro-ph/0105232; D.~J.~Chung, {\it et al.}, Phys.\ Rev.\ D {\bf 64}, 043503 (2001), hep-ph/0104100; A.~Riotto, {\it
In Tegernsee 1999, Beyond the desert 503-521}; K.~Benakli, {\it et al.},
Phys.\ Rev.\ D {\bf 59}, 047301 (1999), hep-ph/9803333; V.~Berezinsky,
M.~Kachelrie, and A.~Vilenkin,
Phys.\ Rev.\ Lett.\  {\bf 79}, 4302 (1997); C.~Coriano, {\it et al.},
Nucl.\ Phys.\ B {\bf 614}, 233 (2001), hep-ph/0107053; S.~Chang, {\it et al.}, Nucl.\ Phys.\ B {\bf 477}, 65 (1996), hep-ph/9605325; H.~Ziaeepour, Astropart.\ Phys.\  {\bf 16}, 101 (2001), astro-ph/0001137.


\bibitem{pro1}
For a review, see F.\,Halzen, Proceedings of the 2001 Lepton-Photon
Symposium, Rome, Italy; R.~A.~Vazquez {\it et al.}, Astroparticle
Physics {\bf 3}, 151 (1995); M.~Ave, {\it et al.}, Phys.\ Rev.\ Lett.\  {\bf 85}, 2244 (2000), astro-ph/0007386.


\bibitem{radio}
P.~Bhattacharjee and G.~Sigl,
Phys.\ Rept.\  {\bf 327}, 109 (2000),astro-ph/9811011; G.~Sigl, S.~Lee,
P.~Bhattacharjee and S.~Yoshida,
Phys.\ Rev.\ D {\bf 59}, 043504 (1999), hep-ph/9809242; R.~J.~Protheroe
and T.~Stanev,
Phys.\ Rev.\ Lett.\  {\bf 77}, 3708 (1996), astro-ph/9605036;
R.~J.~Protheroe and P.~L.~Biermann,
Astropart.\ Phys.\  {\bf 6}, 45 (1996), astro-ph/9605119.


\bibitem{frag}
C.~T.~Hill,
Nucl.\ Phys.\ B {\bf 224}, 469 (1983).

\bibitem{frag2}
S.~Sarkar and R.~Toldra,
Nucl.\ Phys.\ B {\bf 621}, 495 (2002), hep-ph/0108098; C.~Barbot and
M.~Drees, hep-ph/0202072; G.~Sigl, K.~Jedamzik, D.~N.~Schramm and V.~S.~Berezinsky, Phys.\ Rev.\ D {\bf 52}, 6682 (1995), astro-ph/9503094.



\bibitem{agasa}
www.icrr.u-tokyo.ac.jp/as/as.html; www2.keck.hawaii.edu:3636/realpublic/inst
/hires/hires.html.



\bibitem{life}
K.~Hagiwara and Y.~Uehara,
hep-ph/0106320; Y.~Uehara,
hep-ph/0107297; K.~Hamaguchi, {\it et al.}, Phys.\ Rev.\ D {\bf 60}, 125009 (1999), hep-ph/9903207; K.~Hamaguchi, {\it et al.},
Phys.\ Rev.\ D {\bf 59}, 063507 (1999), hep-ph/9809426; K.~Hamaguchi, {\it et al.}, Phys.\ Rev.\ D {\bf 58}, 103503 (1998), hep-ph/9805346.


\bibitem{sim}
I.~F.~Albuquerque, {\it et al.},
Phys.\ Rev.\ D {\bf 64}, 083504 (2001), hep-ph/0009017; A.~E.~Faraggi,
K.~A.~Olive and M.~Pospelov,
Astropart.\ Phys.\  {\bf 13}, 31 (2000), hep-ph/9906345.

\bibitem{spec2}
G.~Jungman and M.~Kamionkowski,
Phys.\ Rev.\ D {\bf 51}, 328 (1995), hep-ph/9407351.


\bibitem{Andres:2001ty}
E.~Andres {\it et al.},
Nature {\bf 410}, 441 (2001).



\bibitem{ice2}
T.~K.~Gaisser, {\it et. al.},
Phys.\ Rept.\  {\bf 258}, 173 (1995), hep-ph/9410384; J.~G.~Learned and
K.~Mannheim,
Ann.\ Rev.\ Nucl.\ Part.\ Sci.\  {\bf 50}, 679 (2000).

\bibitem{gamma}
J.~E.~McEnery {\it et al.}  [Milagro Collaboration],
{\it Prepared for International Symposium on High-Energy Gamma ray
Astronomy, Heidelberg, Germany, 26-30 Jun 2000};
N.~Gehrels  [GLAST Collaboration],
{\it Prepared for International Symposium on High-Energy Gamma ray
Astronomy, Heidelberg, Germany, 26-30 Jun 2000}.



\end{thebibliography}
\end{document}